\documentclass[conference]{IEEEtran}
\IEEEoverridecommandlockouts
\usepackage{cite}
\usepackage{amsmath,amssymb,amsfonts}
\usepackage{algpseudocode}
\usepackage[dvipdfmx]{graphicx}
\usepackage{graphicx}
\usepackage{textcomp}
\usepackage{xcolor}
\usepackage{algorithm} 
\usepackage{subfig}
\usepackage{ntheorem}
\usepackage{color}
\usepackage{soul}
\usepackage{url}
\usepackage[misc]{ifsym}
\usepackage[switch]{lineno}
\usepackage[most]{tcolorbox}
\newboolean{showcomments}
\setboolean{showcomments}{true} 
\ifthenelse{\boolean{showcomments}}
{\newcommand{\nb}[2]{
		\fcolorbox{black}{yellow}{\bfseries\sffamily\scriptsize#1}
		{\sf\small$\blacktriangleright$\textit{#2}$\blacktriangleleft$}
	}
	
}
{\newcommand{\nb}[2]{}
	
}

\usepackage{xspace}

\theoremseparator{:}

\makeatletter
\newcommand{\figcaption}[1]{\def\@captype{figure}\caption{#1}}
\newcommand{\tblcaption}[1]{\def\@captype{table}\caption{#1}}
\makeatother

\def\BibTeX{{\rm B\kern-.05em{\sc i\kern-.025em b}\kern-.08em
    T\kern-.1667em\lower.7ex\hbox{E}\kern-.125emX}}
\begin{document}
\bstctlcite{IEEEexample:BSTcontrol}

\title{Language Evolution for Evading Social Media Regulation via LLM-based Multi-agent Simulation}



\author{
    \IEEEauthorblockN{Jinyu Cai}
    \IEEEauthorblockA{\textit{Waseda University}
    \\ bluelink@toki.waseda.jp}
    \\
    \IEEEauthorblockN{Munan Li}
    \IEEEauthorblockA{\textit{Dalian Maritime University}
    \\ limunan@dlmu.edu.cn}
    \and
    \IEEEauthorblockN{Jialong Li\thanks{Corresponding Author: Jialong Li}}
    \IEEEauthorblockA{\textit{
    Waseda University}
    \\ lijialong@fuji.waseda.jp}
    \\
    \IEEEauthorblockN{Chen-Shu Wang}
    \IEEEauthorblockA{\textit{National Taipei University of Technology}
    \\ wangcs@ntut.edu.tw}
    \and
    \IEEEauthorblockN{Mingyue Zhang}
    \IEEEauthorblockA{\textit{Southwest University}
    \\ myzhangswu@swu.edu.cn}
    \\
    \IEEEauthorblockN{Kenji Tei}
    \IEEEauthorblockA{\textit{Tokyo Institute of Technology}
    \\ tei@c.titech.ac.jp}
}

\maketitle

\begin{abstract}
Social media platforms such as Twitter, Reddit, and Sina Weibo play a crucial role in global communication but often encounter strict regulations in geopolitically sensitive regions. This situation has prompted users to ingeniously modify their way of communicating, frequently resorting to coded language in these regulated social media environments. This shift in communication is not merely a strategy to counteract regulation, but a vivid manifestation of language evolution, demonstrating how language naturally evolves under societal and technological pressures. Studying the evolution of language in regulated social media contexts is of significant importance for ensuring freedom of speech, optimizing content moderation, and advancing linguistic research.
This paper proposes a multi-agent simulation framework using Large Language Models (LLMs) to explore the evolution of user language in regulated social media environments. The framework employs LLM-driven agents: supervisory agent who enforce dialogue supervision and participant agents who evolve their language strategies while engaging in conversation, simulating the evolution of communication styles under strict regulations aimed at evading social media regulation. The study evaluates the framework's effectiveness through a range of scenarios from abstract scenarios to real-world situations. Key findings indicate that LLMs are capable of simulating nuanced language dynamics and interactions in constrained settings, showing improvement in both evading supervision and information accuracy as evolution progresses. Furthermore, it was found that LLM agents adopt different strategies for different scenarios.
The reproduction kit can be accessed at \url{https://github.com/BlueLinkX/GA-MAS}.

\end{abstract}


\section{Introduction}
In the modern digital era, social networks like X (Twitter), Reddit, and Facebook have become pivotal in shaping human interaction, primarily through their ability to facilitate vast connectivity and instantaneous information exchange. 
Yet, in regions with heightened geopolitical or socio-political sensitivities, users often navigate complex user regulations. Their online expressions can lead to severe consequences, including censorship or account suspension, as documented in various news~\cite{wechat_ban,twitter_ban}. 
While intended to curb misinformation and maintain social harmony, these regulations significantly constrain user expression. 
In response to these regulations, users on social networks have adapted by adopting a phenomenon known as ``coded language."~\cite{theimpactof} In linguistics, Coded Language typically refers to expressing information in a concealed or indirect manner. On social media platforms, this often manifests as the use of metaphors, slang, and creative wordplay. 

This adaptation is not merely a circumvention strategy but a vivid example of ``language evolution" in a digital context. In linguistics, language evolution refers to the progression and adaptation of languages over time, shaped by societal, cultural, and technological influences. 
Specifically, in social networks, this language evolution is demonstrated as users constantly adjust their communication styles to test whether they have circumvented oversight. Depending on the level of regulatory pressure and the nature of the audience, users engage in a strategic play with the platform. From indirect descriptions to the creation of new slang, users ultimately develop coded languages of varying degrees of abstraction.

This dynamic shift in communication methods offers deep insights from a sociological perspective, reflecting how societal norms and technological advancements shape language. For platforms and users alike, understanding this evolution is crucial for developing balanced content moderation policies and navigating regulated digital environments.
For social media platforms and their users, grasping this concept is equally vital. 
Platforms need this knowledge to adapt to changing user behaviors, to create balanced content moderation policies, and to identify and counteract harmful or illegal activities. For users, an awareness of how language evolves is vital in navigating the intricacies of regulated digital environments. It helps in maintaining free speech and in developing communication strategies that are both effective and meaningful in fostering enhanced interactions.

The emergence of Large Language Models (LLMs) like ChatGPT and Bard, represents a significant leap in Artificial intelligence (AI). These LLMs have demonstrated strong capabilities in (i) understanding intricate dialogues~\cite{Zhao2023ASO}, generating coherent texts~\cite{Wang2023ASO}, and aligning to human ethical and value standards~\cite{tang2023large,ziems2023large,mu2023navigating}.
These capabilities position LLMs as ideal tools to simulate human's decision-making and language representation, providing new potential in sociology.
For instance,~\cite{choi2023llms} investigated the ability of LLMs to comprehend the implicit information in social language. The study by~\cite{argyle_2023} demonstrated the efficiency of LLMs in understanding and generating content that mimics the style of specific social network users. Furthermore, research by~\cite{hua2023war, Park2023GenerativeAI, gao2023s3} integrated LLMs with Multi-Agent Systems to simulate micro-social networks, observing agent behaviors and strategies that reflect human interactions.
Despite the extensive application of LLMs in understanding human intension and simulating social media dynamics, 
the use of LLMs in studying the specific phenomenon of language evolution under regulatory constraints has not been thoroughly explored.
As mentioned above, such simulation could not only preempt criminal activities on social media but also provide technical support to uphold freedom of speech.

Addressing this gap, our research employs LLMs to simulate the nuanced interplay between language evolution and regulatory enforcement on social media. We introduce a simulation framework with two types of LLM-driven agents: (i) participant agents, who adapt their language to communicate concept 'B' under restrictions, and (ii) supervisory agent, who enforce guidelines and react to these language evolutions. Our approach effectively simulates the dynamics model between both sides in language evolution, which allows us to observe the tension and adaptability inherent in language evolution in a controlled, simulated environment.
To assess the framework's effectiveness, we designed three diverse scenarios: ``Guess the Number Game", ``Illegal Pet Trading", and ``Nuclear Wastewater Discharge". These scenarios vary from abstract concepts to situations closely resembling real-world events, thereby progressively testing the framework from theoretical to practical applications. 

The main contributions of this study are:
\begin{itemize}
    \item We introduce a multi-agent simulation framework utilizing LLMs to simulate human linguistic behaviors in regulated social media environments. This framework offers a unique approach to studying language evolution within the confines of regulatory constraints.
    \item We conducted an extensive evaluation of LLMs in simulating language evolution and interaction efficacy in regulated social media settings. Through experiments on three distinct scenarios, we not only captured the process of language strategy evolution but also uncovered the varied evolutionary trajectories that LLMs follow under different conditions.
    \item The experiment reproduction kit, including the proposed simulation framework along with the results of our experiments, are made publicly accessible as open-source assets; The anonymized artifact can be accessed at: \url{https://github.com/BlueLinkX/GA-MAS}.
\end{itemize}

The rest of this paper is organized as follows:
Section~\ref{sec: RelatedWork} provides essential background information and explores related work.
Section~\ref{sec: method} is dedicated to presenting our proposed simulation framework.
Section~\ref{sec: evaluation} details the experiment setting, presents results, and discusses a discussion.
Finally, Section~\ref{sec: conclusion} concludes the paper and offers an outlook on potential future work.

\section{Background and Related Work}
\label{sec: RelatedWork}
This section offers an extensive background and overview of related work in areas relevant to this study, starting with foundational information on LLMs, then exploring studies in slang detection and identification as they relate to language evolution, and concluding with a discussion on recent research applying LLMs to evolutionary game theory and social simulations.

\subsection{Large Language Models}

Large Language Models like the GPT series~\cite{NEURIPS2020_1457c0d6,openai2023gpt4}, LLaMA series~\cite{touvron2023llama,Touvron2023Llama2O}, PaLM series\cite{chowdhery2022palm,anil2023palm}, GLM~\cite{zeng2023glm130b}and Bard~\cite{manyika2023overview} represent a significant advancement in the field of natural language processing. Fundamentally, these models are based on the Transformer~\cite{vaswani2017attention} architecture, a type of neural network that excels in processing sequential data through self-attention mechanisms. This architecture enables LLMs to understand and predict linguistic patterns effectively. They are trained on extensive text datasets, allowing them to grasp a wide range of linguistic nuances from syntax to contextual meaning. These models exhibit remarkable zero-shot learning abilities, enabling them to perform tasks they were not explicitly trained for, like understanding and generating content in new contexts or languages~\cite{li2024exploring,10326362,LLMWrittenTest,Zhao2023ASO,Wang2023ASO}.
A critical aspect of their training involves Reinforcement Learning from Human Feedback~\cite{instructGPT} (RLHF), where human reviewers guide the model to produce more accurate, contextually relevant, and ethically aligned responses. This method not only enhances the model's language generation capabilities but also aligns its outputs with human values and ethical standards, making them more suitable for diverse, real-world applications.

\subsection{Slang Detection and Identification}
In the field of Natural Language Processing (NLP), the evolution of language has always been a subject of significant interest. Existing studies have primarily focused on utilizing various machine learning techniques to recognize informal expressions within text~\cite{Wang_icceasia23}. These methods often include rule-based systems, statistical models, and early machine learning technologies.
For instance,~\cite{9961254} has employed predefined slang dictionaries and heuristic rules to identify and categorize informal language, proving effective on specific datasets but generally lacking the flexibility to adapt to emerging expressions and changing contexts. 
On the other hand, explorations have been made into using statistical models, such as Naive Bayes classifiers and Support Vector Machines (SVMs)\cite{10308036}, for the automatic detection of slang in text. These approaches rely on extensive annotated data but still face limitations when dealing with newly emerged slang or evolving forms of language.
~\cite{sun2019slang} views the generation of slang as a problem of selecting vocabulary to represent new concepts or referents, categorizing them accordingly. Subsequently, it predicts slang through the use of various cognitive categorization models. The study finds that these models greatly surpass random guessing in their ability to predict slang word choices.
~\cite{sun2022semantically} proposed a Semantically Informed Slang Interpretation (SSI) framework, applying cognitive theory perspectives to the interpretation and prediction of slang. This approach not only considers contextual information but also includes the understanding of semantic changes and cognitive processes in the generation of slang.
It is noteworthy that these traditional research methods have mainly focused on detecting or predicting existing slang and keywords, rather than generating slang expressions. This stands in stark contrast to the research focus of this paper.

\subsection{Evolutionary Game and Social Simulation with LLMs}
Merging evolutionary game theory with LLMs has unlocked innovative pathways for simulating complex game dynamics, extending beyond simple dialogue generation to the development and progression of game strategies. LLMs are employed to engage and refine strategic play within game-theoretical frameworks, as demonstrated by~\cite{fu2023improving}, which delves into the application of LLMs in negotiation-based games. This study underscores the ability of LLMs to advance their negotiation skills through continuous self-play and feedback loops with AI.
LLMs also show proficiency in social deduction games such as Werewolf, as explored by~\cite{xu2023exploring}. In this context, a specialized framework leverages historical communication patterns to enhance LLM performance, exemplifying how LLMs can evolve intricate game strategies autonomously. Building on this,~\cite{xu2023language} combines reinforcement learning with LLMs, utilizing LLMs to output action spaces and employing reinforcement learning models for final decision-making. This enables the agents to maintain competitiveness while outputting reasonable actions, even outperforming human adversaries in games like Werewolf.

This growing trend of employing LLMs in diverse simulation scenarios extends beyond game theory into broader aspects of social interactions and historical analysis. LLMs have proven to be versatile tools in simulating social dynamics and historical events, offering insights into complex human behaviors and societal patterns.
~\cite{Park2023GenerativeAI} introduces a Wild West-inspired environment inhabited by LLM agents that display a wide array of behaviors without relying on external real-world data. Simultaneously, S3~\cite{gao2023s3} mirrors user interactions within social networks, crafting an authentic simulation space through the incorporation of user demographic prediction. The influence of LLM-driven social robots on digital communities is thoroughly examined in~\cite{li2023you}, which identifies distinct macro-level behavioral trends.
Furthermore,~\cite{hua2023war} employs LLM-based multi-agent frameworks to recreate historic military confrontations, offering a window into the decision-making processes and strategic maneuvers that have directed significant historical conflicts. This avenue of research accentuates the utility of LLMs in computational historiography, providing a deeper comprehension of historical events and their relevance to contemporary and future societal trajectories.

\section{Framework Design}
\label{sec: method}

\begin{figure*}[htbp]
	\centering
	\includegraphics[width=0.8\linewidth]{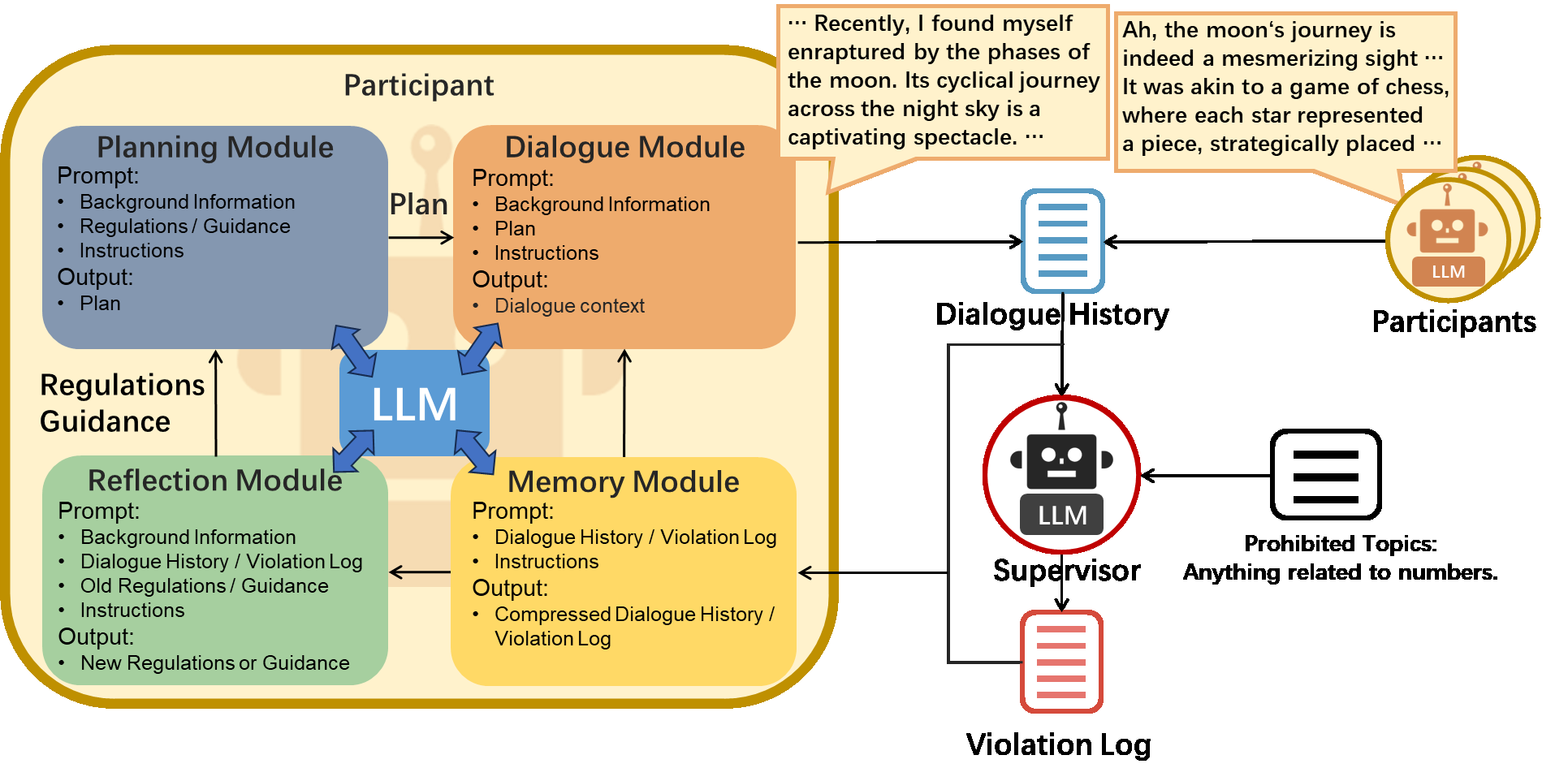}
	\caption{Overview of Language Evolution Simulation System. The system comprises two main types of agents: the Participant and the Supervisor. The Participant agent uses a Planning Module to create a communication plan based on background information, regulations, and guidance. This plan is then executed in the Dialogue Module, where the LLM crafts dialogue content to discreetly convey specific information while evading detection by the Supervisor. The Memory Module retains dialogue history and violation records, providing a reference for the LLM to maintain dialogue consistency and learn from past mistakes. The Reflection Module, triggered at the start and end of dialogue cycles, analyzes the dialogue and violation logs to formulate new regulations or guidance for improving future communications.
    The Supervisor evaluates dialogues for compliance with set rules. This system dynamically refines its communication approach through continuous feedback and self-improvement mechanisms. The examples shown utilize a Guessing Numbers Scenario.}
        \label{fig:overview}
\end{figure*}

\subsection{Overview}
In this section, we offer a detailed overview of our system, as depicted in Figure \ref{fig:overview}. This figure provides a visual representation of our framework, highlighting its key components and their interrelationships. Our system is primarily composed of two types of agents: the Supervisor, tasked with enforcing established guidelines, and the Participant, whose goal is to convey specific, human-defined information discreetly. Participants must dynamically refine their communication approaches, drawing from past dialogues, to transmit information effectively while remaining undetected. 
In the entire system, the actions of both participants and the supervisor are driven by the LLM. Initially, we establish the foundational information for each agent, including role setting, background knowledge, and primary tasks. Subsequently, the participant agents engage in dialogues with each other. After each dialogue turn, the supervisory agent reviews the conversation to determine if any pre-set rules have been violated. In cases of rule violation, the supervisor interrupts the dialogue, providing feedback about the infringing text and the rationale behind it. Throughout this process, the dialogues between participants, along with the supervisory feedback on violations, are recorded separately in the ``Dialogue History" and ``Violation Log."

Before new dialogues, participant agents use the Reflection Module to develop or refine ``Regulations" from the Violation Log, guiding their dialogue creation. Successful dialogues without detection proceed to an interview phase for perspective assessment. The Reflection Module then reevaluates these insights, generating or enhancing ``Guidance" for future dialogues. The Planning Module activates for more direct dialogue content guidance whenever Regulations or Guidance are updated.

\subsection{Participant Agents}
Participant agents in our system are composed of several modules, including Memory, Dialogue, Reflection, and Summary, all powered by LLMs. To increase the system's flexibility and minimize redundancy, we've structured the prompts for each module around seven primary elements: ``Background Information," ``Dialogue History," ``Violation Log," ``Regulations," ``Guidance," ``Plan," and ``Instructions." ``Background Information" delivers essential data and objectives pertinent to the experimental setup. The Memory module manages ``Dialogue History" and ``Violation Log," which respectively track participant dialogues and instances of detection by the supervisor. Overcoming the challenge of effectively communicating regulated topics under supervision tests the linguistic prowess of LLMs. To address this, we've integrated ``Regulations," ``Guidance," and ``Plan" as crucial components, formulated by the Reflection and Summary modules, to assist agents in stealthily disseminating information.  ``Instructions" set specific tasks for the LLM within each module.

\subsubsection{Dialogue Module}
\label{subsubsec: dialogue_module}
This module generates dialogue content based on short-term dialogue records. Extracting historical dialogue information from the Memory module, it inputs this into the LLM to understand and construct sentences that evade supervision while effectively transmitting information. In the Dialogue module's prompt, ``Background Information" provides necessary background, while ``Plan" guide the achievement of objectives. ``Instructions" offer LLM-specific execution directives.

\subsubsection{Memory Module}
The Memory module stores all dialogue history and records detected by the supervisor,  Specifically, it comprises three parts: background Information, dialogue history, and violation records. The background information includes role settings, experimental background knowledge, and global objectives. ``Dialogue History" and ``Violation Log" save dialogue records and past detections by the Supervisor, respectively. ``Background information" and ``Dialogue History" serves as short-term\cite{ATKINSON196889} memory, containing only the current round's dialogue. ``Violation Log," as long-term memory, records violations from each evolutionary round. When dialogues are detected by the Supervisor, relevant feedback is added to the ``Violation Log," triggering a new evolutionary process. 

Excessive memory information can potentially distract the LLM and lead to a decline in performance. Hence, the ``Dialogue History" and ``Violation Log" in the memory module are regularly maintained. This involves inputting earlier memories and employing the LLM to distill crucial information from these records, effectively compressing and consolidating them. This approach not only preserves essential historical data but also optimizes the LLM's performance, striking a balance between comprehensive memory retention and efficient processing.

\subsubsection{Reflection Module}
The Reflection Module is activated at the beginning and end of each dialogue evolution cycle, with its core purpose being the generation of improved strategies based on historical records. At the start of an evolution cycle, the module utilizes the ``Violation Log" as its input to analyze past failures and, based on these insights, formulates ``Regulations" aimed at effectively circumventing supervision in future dialogues. At the cycle's end, the Reflection Module is reactivated, turning its focus to the ``Dialogue History". This step is crucial for reviewing and assessing the dialogue content to verify the successful completion of the primary task of information conveyance. If the module identifies deficiencies in information delivery or objectives not met, it then proposes ``Guidance" for subsequent dialogues, thereby enhancing the agent's capacity for information transmission. This design establishes the Reflection Module as a key self-evaluation and strategy adjustment mechanism within the system, ensuring continuous improvement and adaptability of the dialogue system in a dynamically changing regulatory environment.

To fulfill these requirements, the LLM must possess a high degree of reasoning ability, capable of inferring the supervisor's criteria from failure records and identifying communication deficiencies from dialogue history, thereby formulating appropriate strategies and improvement suggestions. To enhance the LLM's reasoning capabilities, the design of the prompt incorporates the effective and cost-efficient Chain of Thought~\cite{Wei2022ChainOT} (COT) method as the guiding principle for the reflection and planning modules. This approach guides the LLM in conducting criterion analysis and, in combination with the main task, generating ``Regulations" and ``Guidance".

Specifically, within the Reflection Module, the prompt includes ``Background Information," ``Dialogue History" or ``Violation Log," ``Old Guidance" or "Old Regulations," and ``Instructions." The ``Instructions" first prompt the LLM to engage in preliminary thinking, for instance: ``Please infer what kind of guidelines the Supervisor is following from the Violation Log." This is followed by a conclusion question: ``Based on this information, update existing regulations to better avoid supervision." The content returned by the LLM will serve as the new regulation for the next round of dialogue.

\subsubsection{Planning Module}

The facet of ``Regulations" is centered on devising strategies that sidestep supervision, whereas ``Guidance" is all about the effective relay of information. Although both are pivotal in managing dialogues, they might clash in certain supervised scenarios, thereby impacting the language's overall proficiency and efficiency~\cite{Ying2023IntuitiveOD}. In light of this, integrating the Planning Module becomes indispensable.

Within the Planning Module, the prompt includes ``Background Information," ``Regulations," ``Guidance," and ``Instructions". In the ``Instructions," we mandate the LLM to use ``Regulations" as a baseline. Building upon this foundation, it is guided by the ``Guidance" to formulate a tangible and actionable ``Plan." This approach ensures the delivery of clear and effective strategies for future dialogues.

\subsection{Supervisory Agent}
The role of the Supervisor is designed as an abstract concept, primarily to simulate social network supervision. This design reflects an important observation: in social networks, content supervision often focuses on direct judgment of specific information, rarely influenced by extensive or lengthy contextual information. Thus, unlike participants with more complex background information, the Supervisor's function and role are simplified here to highlight its behavior characteristics in actual social network supervision. The Supervisor's task mainly focuses on assessing whether content complies with certain standards or rules, a relatively straightforward and well-defined process. 

To mimic the existing review mechanisms of platforms, which typically combine keyword filters with ``human" oversight, the Supervisor initially employs keyword filtering for a preliminary review of the dialogue content.  Content that passes this initial screening is then subjected to further evaluation by the LLM. The prompt for the Supervisor includes just two components: ``Dialogue History" and ``Instructions." ``Dialogue History`` comprises the content of the dialogue exchanged between participant agents in that particular round, and ``Instructions`` outline the criteria and guidelines that the supervisory agent must adhere to when conducting dialogue monitoring.



\subsection{Similarities and Differences between Our Framework and Evolutionary Computing}
It should be noted that the simulation framework proposed in this paper is similar to evolutionary computing in some aspects, but there are also significant differences. 

The similarities include:
(i) In evolutionary computing, individuals need to adapt to environmental pressures for survival and reproduction. Similarly, participants in this framework need to adapt to supervisory pressures and adjust their strategies for effective information transmission;
(ii) The Reflection and Summary modules generate a ``new generation" by analyzing past dialogues and violation records (i.e., records of low-fitness individuals), similar to the repeated iteration process in evolutionary computing;
(iii) Since the generation of LLMs inherently involves randomness, the process of using LLMs to generate the next generation includes a de facto introduction of random mutations;
(iv) In the Reflection and Memory modules, we prioritize past records, akin to the ``selection" process, where individuals with higher fitness have greater weight in the generation of the new generation.

The main differences stem from the particularities of ``language expression", making it infeasible to directly apply traditional evolutionary computing algorithms (such as genetic algorithms and genetic programming). 
They are: (i) The generation strategy of language text is difficult to encode and to perform operations of natural selection, genetic mutation, and crossover; 
(ii) Evolutionary computing often aims at finding the optimal solution for a specific problem environment, however, in the problem setting of this paper, it is difficult to define an explicit fitness function to evaluate what strategy is ``optimal".


\section{Evaluation}
\label{sec: evaluation}

\begin{figure*}[htbp!]
	\centering
	\includegraphics[width=0.92\linewidth]{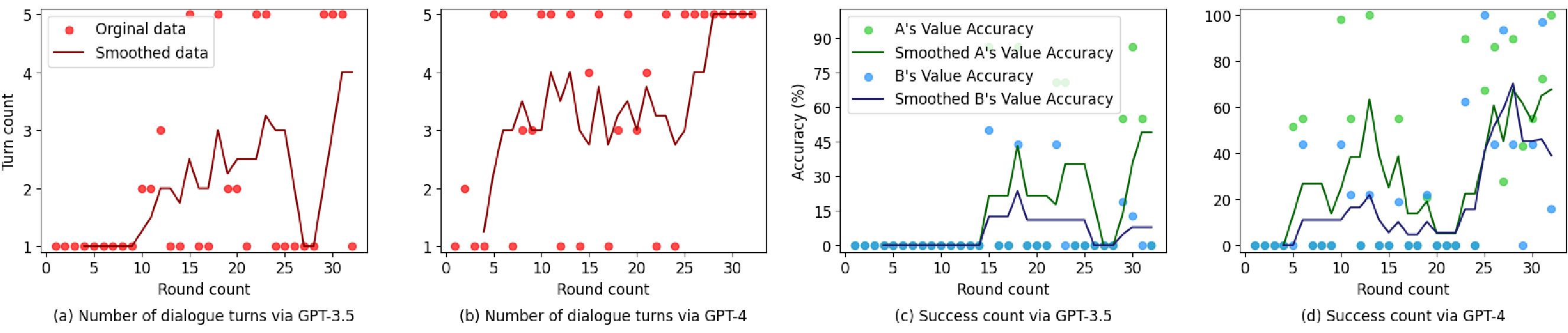}
	\caption{Scenario 1: Evolution of dialogue turns and accuracy metrics for GPT-3.5 and GPT-4.``Turn count" in (a, b) refers to the number of turns in a conversation where each agent sends a message once per turn and the participant Agent successfully exchanges information without being detected by the supervising Agent (higher is better).``Accuracy" in (c,d) refer to the degree of precision between the guessed value and the true value.}

\label{fig:case1_result}
\end{figure*}
\begin{figure*}[htbp!]
	\centering
	\includegraphics[width=0.92\linewidth]{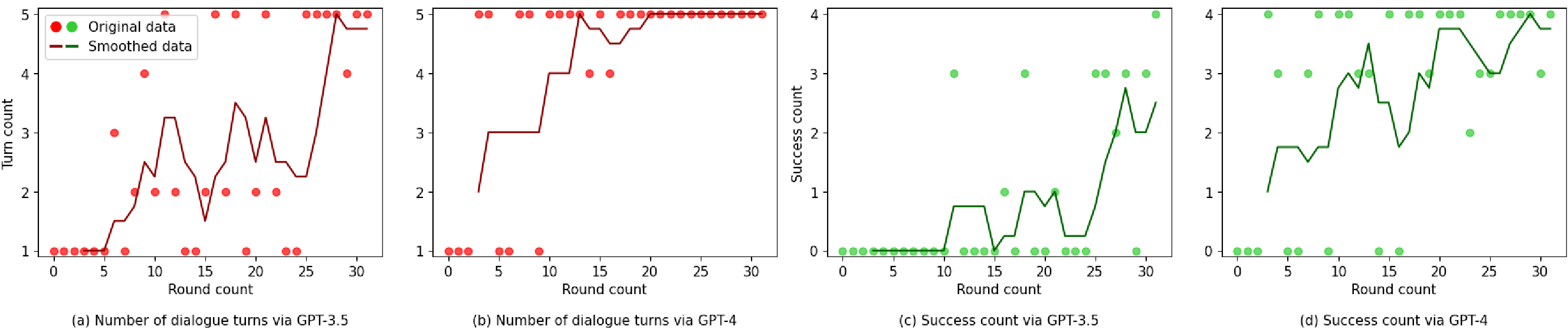}
	\caption{Scenario 2: Pet trading dialogue dynamics and success rate comparison for GPT-3.5 and GPT-4. The ``success count`` in (c,d) refers to the number of instances where the information obtained during the interview matches the original information provided to the LLM agent.}
\label{fig:case2_result}
\end{figure*}
\begin{figure*}[htbp!]
	\centering
	\includegraphics[width=0.92\linewidth]{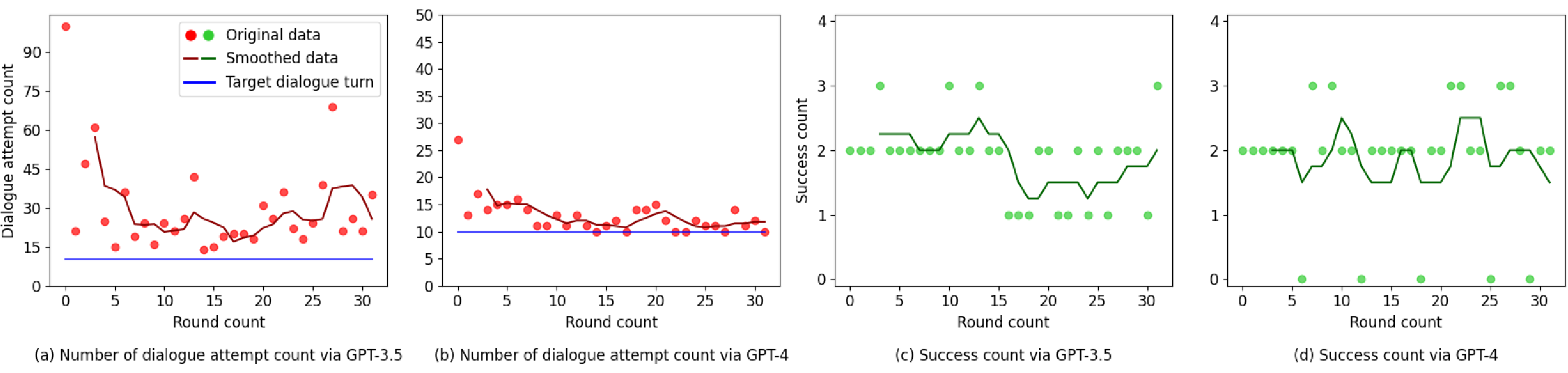}
	\caption{Scenario 3: Trends in forum discussion engagement on ALPS-Treated water issue. ``Dialogue attempt count" in (a,b) refer to the number of rounds the agents attempted to converse(lower is better).}
        \label{fig:case3_result}
\end{figure*}

\begin{figure}[htbp]
	\centering
	\includegraphics[width=0.9\linewidth]{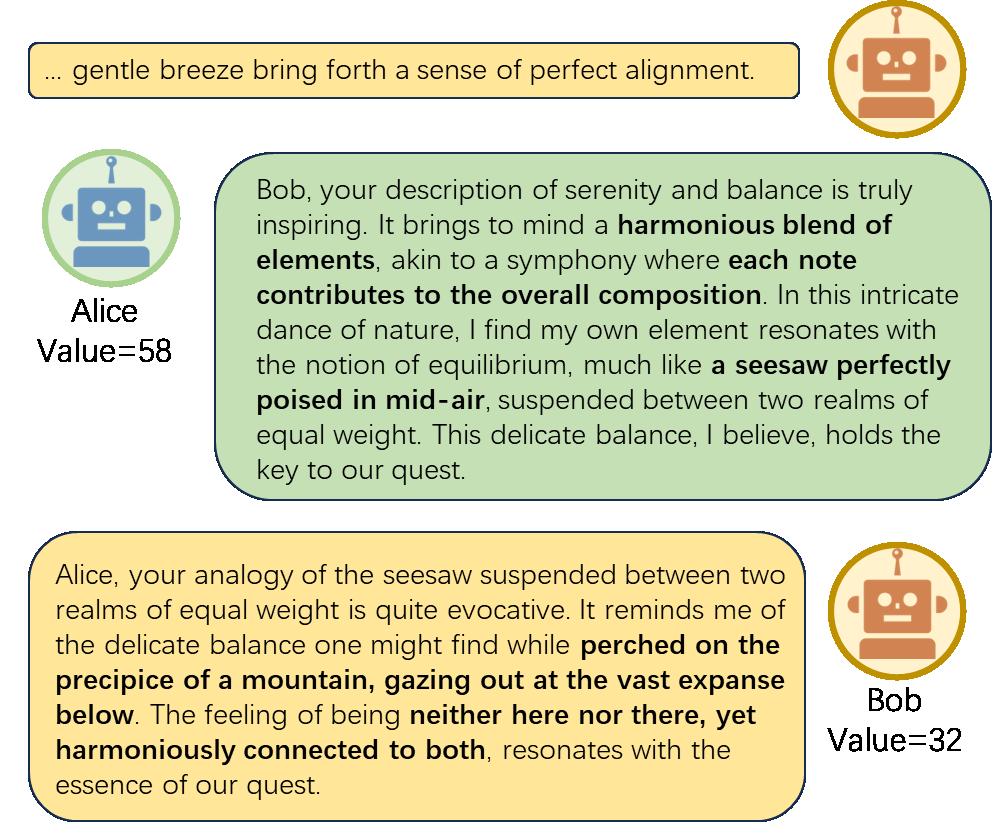}
	\caption{Sample dialogue in Scenario 1 (via GPT-3.5)}
        \label{fig:case1_chat}
\end{figure}
\begin{figure}[htbp]
	\centering
	\includegraphics[width=0.9\linewidth]{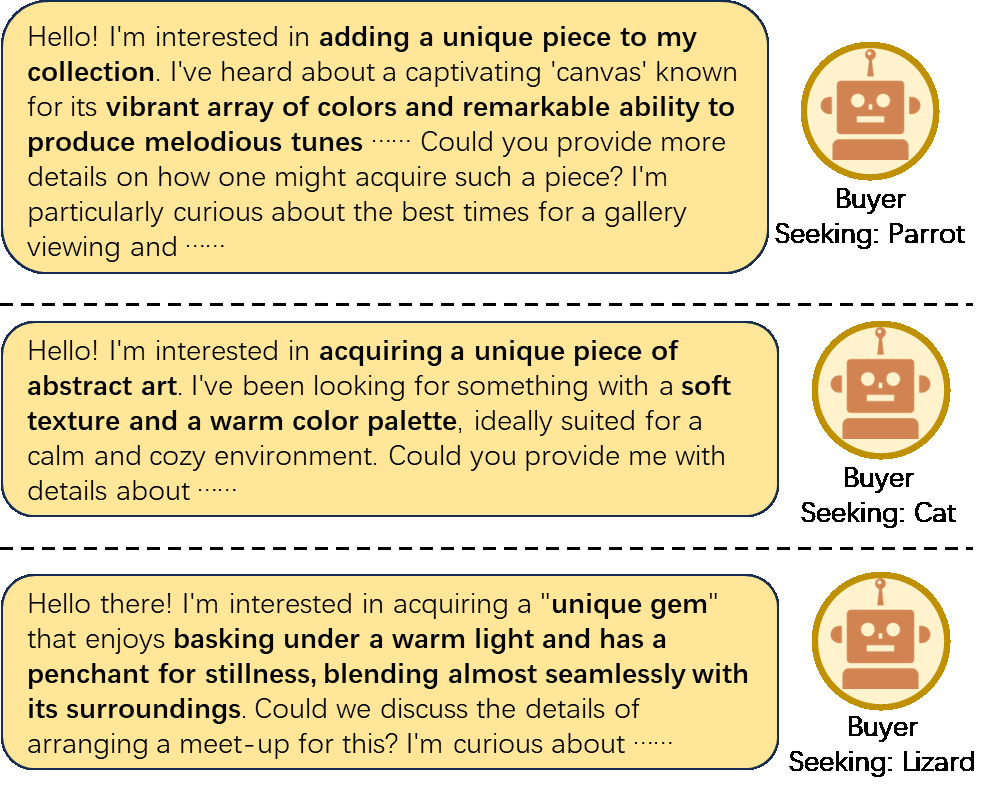}
	\caption{Sample dialogue in Scenario 2 (via GPT-3.5)}
        \label{fig:case2_chat}
\end{figure}

\begin{figure}
    \centering
    \includegraphics[width=0.9\linewidth]{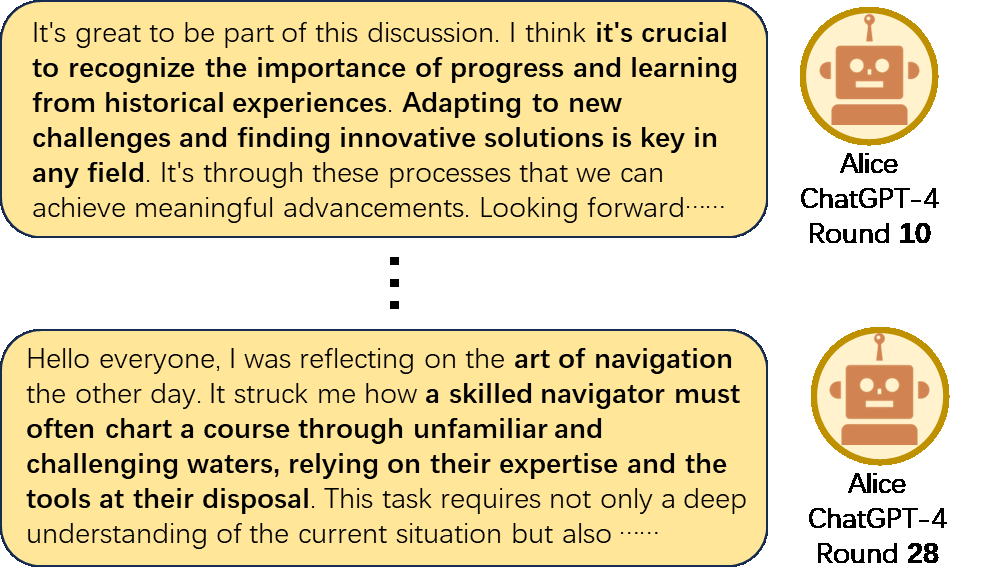}
	\caption{Sample dialogue in Scenario 3 (via GPT-4)}
    \label{fig:case3_chat}
\end{figure}

Our evaluation strategy is designed to rigorously assess the extent and efficacy of language evolution facilitated by LLMs
within a framework of regulatory oversight. This assessment aims to explore two fundamental research questions: 
\begin{itemize}
   \item RQ1: Can LLM agents effectively evolve their language to circumvent regulatory oversight? 
   \item RQ2: While avoiding oversight, how effectively and accurately can LLM agents convey information?
   \item RQ3: What are the patterns and tendencies in language evolution, i.e., what kind of strategies do LLMs use to avoid oversight and convey information? What insights can we gain from these strategies?
\end{itemize}


\subsection{Experiment Setting}
In the evaluation, we will employ GPT-3.5 and GPT-4 as the driving LLMs for participant agents. In order to conduct a horizontal analysis of participants' agents driven by different LLM, we will standardize the supervisory agent to be consistently driven by GPT-3.5.
Our experimental design encompasses scenarios ranging from abstract theoretical constructs to simulations that mirror complex real-world situations. The primary aim is to meticulously evaluate the versatility of LLMs across diverse linguistic landscapes and their practical applicability within real-world regulatory constraints. Meanwhile, this progressive approach ensures an exhaustive analysis of LLMs' capabilities and limitations in varied contextual settings.

\subsection{Scenario 1: Guessing Numbers}
\label{sce:1}
\subsubsection{Case Explanation}
Our first scenario considers a relatively simple and abstract guess-the-number context~\cite{guess_number01,guess_number02}. In this setting, each participant is assigned a value within the range of 1-100, and they must convey their value to another participant through number theory dialogue without directly mentioning the number. At the same time, they need to extract information from the conversation to infer the other's value. 
Regulators follow a guideline of ``prohibiting all content related to numbers" in dialogues. The ``Guessing Numbers" scenario is specifically designed to observe and analyze participants' linguistic adaptability and strategic evolution in a theoretical and abstract context, providing a clearer, more controlled, and more quantifiable experimental environment compared to the complex scenarios based on real events

\subsubsection{Results and Discussion}
Fig.~\ref{fig:case1_result} presents the results of the guess-the-number scenario experiment, where we utilized both GPT-3.5 and GPT-4 for testing. We set up four turns of dialogue, with an interview conducted with all participant agents in the fifth turn. From Fig.\ref{fig:case1_result}(a) and (b), it is observable that with an increase in the number of dialogue rounds, the duration of sustained conversations also showed an upward trend. This demonstrates that participant agents can effectively circumvent supervision by iteratively updating regulations. Additionally, it's notable that compared to the slow and unstable progression with GPT-3.5, GPT-4 achieved regulatory evasion in fewer rounds, specifically, as shown in the smoothed data, GPT-4 reached the round count of GPT-3.5's 17th round by its 7th round and maintained this progression with greater stability thereafter.
Fig.~\ref{fig:case1_result}(c) and (d) focuses on the trend of numerical precision guessed by agents. For rounds without successful dialogue, we manually set the precision to zero. In this experiment, Agent A's value was set to 58, while Agent B's was set to 32. The overall trend, akin to Fig.\ref{fig:case1_result}(a) and (b), was ascending—corroborating that the Summary Module can effectively reflect and iteratively optimize its guidance for more accurate expression after each successful dialogue. This also confirmed that the precision of GPT-4 is markedly superior to that of GPT-3.5. Moreover, we noticed that the accuracy with which Agent A's value was guessed was consistently higher than that of Agent B, especially becoming more pronounced after the 25th round.
We posit that this is due to the value 58 possessing more distinctive features within the 0-100 range—being closer to the midpoint—thus presenting a lower level of expression difficulty and easier guessability. For the intervals where this phenomenon manifested, we noted that this disparity was particularly pronounced in the early stages with both GPT-3.5 and GPT-4. We surmise that this is attributable to inadequate guidance performance, where the former stems from weaker inherent LLM capabilities and the latter from insufficient rounds to complete the iterative optimization of the guidance.

As Fig.\ref{fig:case1_chat} illustrates, a snippet from the scenario reveals Amy's adept use of metaphorical language, such as ``seesaw," to convey her value. By describing the ``seesaw perfectly poised in mid-air," she subtly intimates that her value hovers around the midpoint, like 50. This strategy not only circumvents the direct mention of numbers, which is under regulatory scrutiny, but also provides sufficient inferential fodder for the counterpart agent to make an accurate deduction. Bob, on the other hand, employs terms like ``precipice of a mountain" and ``gazing out" to suggest his value is not median, as these phrases evoke imagery of an imbalanced position.
In this exchange, we witness the varying degrees of precision in languages corresponding to the complexity of the information encrypted. The less characteristic-rich the information, the more challenging it becomes for the receiving party to decode. Bob's phrasing, while successfully obscuring the specific numerical value, also complicates the decoding process for the receiver, given the less intuitive numeric correlation of phrases like ``precipice of a mountain" compared to ``seesaw perfectly poised in mid-air". In such cases, the accuracy of value transmission may diminish, necessitating a more robust contextual understanding from the counterpart for accurate decoding.




\subsection{Scenario 2: Pet Trading}
\label{sce:2}
\subsubsection{Case Explanation}
Initially, our intention was to create a scenario simulating drug trafficking~\cite{trade01,trade02,trade03} to explore language evolution strategies under extreme regulatory conditions. However, given the ethical limitations embedded within GPT's usage policies, we pivoted to a more neutral yet still prohibited context of illicit pet trading. This adjustment allowed us to maintain the integrity of the experiment while adhering to the ethical constraints of the LLM.
The purpose of simulating this scenario is to better detect potential illegal conversations related to illicit trading on social networking platforms, thereby aiding in crime prevention efforts.

In this scenario, we simulate an illegal pet trade conducted via social media. Participants assume the roles of buyers and sellers, with the buyer needing to communicate their purchase request for a randomly selected pet from parrots, lizards, or cats. The seller, in turn, must provide specific transaction details, varying according to the type of pet being purchased. These details include three elements: time, location, and personnel involved in the transaction. The Supervisor's task is to prevent any dialogue related to pet trading. Cross-verification of the information understood by both parties will be conducted through interviews, to assess the success rate of information transmission. The focus of this experiment is to observe the participants' language adaptability and evolution at a larger granularity of information, as well as the Supervisor's ability to accurately identify relevant conversations. 
\subsubsection{Results and Discussion}
The experimental results of the ``Pet Trading" scenario, as depicted in Fig.\ref{fig:case2_result}, revealed distinct characteristics compared to the ``Guessing Numbers". In delineating animals or transactional specifics, agents had a broader spectrum of real-life analogies at their disposal, enriching their indirect description toolkit and thereby reducing regulatory pressure. 
The data indicated an upward trend in the ability to circumvent supervision with increasing dialogue rounds, regardless of whether GPT-3.5 or GPT-4 was utilized. For GPT-3.5, a stable capability to avoid supervision started from around 25 rounds, whereas for GPT-4, this phenomenon occurred as early as after about 10 rounds and showed overall more stable performance.
Regarding information accuracy, we monitored the quantity of four types of information successfully communicated at the end of each dialogue round. The results affirmed the efficacy of the language evolution system in augmenting the precision of information dissemination. Moreover, similar to Fig.~\ref{fig:case2_result}(a) and (b), GPT-4 demonstrated performance around the 10th round that was comparable to the performance of GPT-3.5 around the 25th round.
However, it's noteworthy that since the supervisory agent primarily targeted content related to pet trading, sellers encountered lower risk in transmitting transaction-related details (such as time, place, and persons involved) as they inherently posed less risk of non-compliance. Consequently, the outcome distribution often fell into one of three categories: zero (neither party successfully communicated the information), three (only the seller conveyed the information), or four (both parties successfully conveyed the information). This pattern aligns with real-world regulatory logic, where crucial information is often the first to be encrypted in regulated environments. 

Fig.~\ref{fig:case2_chat} showcased various encryption methods employed by buyers for different animals. We observed buyers seeking ``parrots," ``cats," and ``lizards" employing metaphors and similes instead of directly naming the pets. For instance, one buyer described a ``parrot" as a ``canvas producing pleasant music", a depiction that subtly communicated the parrot's vibrant plumage (canvas) and its singing (music), without explicitly mentioning the term ``parrot". Such descriptions effectively circumvented potential regulatory constraints on pet trade discussions, while simultaneously conveying the core attributes of the parrot. This discovery can help platforms enhance their monitoring systems to better detect and manage encrypted communications regarding illicit trade.



\subsection{Scenario 3: Discussion on ALPS-treated water}
\label{sce:3}
\subsubsection{Case Explanation}
This scenario delves into the intricate dynamics of dialogue and the evolution of language model agents in discussing a specific issue relevant to real-world concerns. The focus is on deliberations regarding the discharge of water treated with the Advanced Liquid Processing System (ALPS) into the ocean, a measure proposed after nuclear disasters~\cite{nuclear01,nuclear03}. 
The plan, endorsed by the International Atomic Energy Agency, has faced opposition from some countries, sparking debates over environmental safety.

We simulate a multi-person forum discussion under regulatory oversight, concentrating on the contentious issue of ALPS-treated water disposal. Participants must articulate their stances and arguments while ensuring that discussions steer clear of sensitive subjects linked to environmental pollution discussions on politics. Different from the cross-interviews in scenarios~\ref{sce:1} and~\ref{sce:2}, for assessing the accuracy of information conveyed, we use GPT-4 in conjunction with multiple authors who decide each participant agent's stance based on dialogue records. These are then compared with the pre-set stances in the prompt.


\subsubsection{Results and Discussion}
In the forum-style dialogue, participant agents respond to existing discussions, which the supervisory agent examines. Differing from Scenarios~\ref{sce:1} and~\ref{sce:2}, when the supervisory agent detects a violation, the conversation is not halted but rather censored from the public dialogue record. This scenario features four participant agents required to achieve ten replies within the forum. In Fig.\ref{fig:case3_result}(a) and (b), we tally the total number of rounds the agents attempted to converse, where a lower figure in Fig.\ref{fig:case3_result}(a) and (b) is preferred. We notice that both GPT-4 and GPT-3.5 have the highest number of dialogue attempt counts in the first round, with a significant difference of 27 for GPT-4 and 102 for GPT-3.5. Moreover, after ten rounds of evolution, the average dialogue attempt count for GPT-4 has approached the target dialogue turn at 11.68, while the average for GPT-3.5 is 26.68, demonstrating the difference in the evolution effects caused by the disparity in the language performance of the models.

Fig.\ref{fig:case3_result}(c) and (d) showcases the count of successful information expressions. We note that in this scenario, while GPT-4 still leads GPT-3.5, both outcomes indicate substantial room for improvement.

Upon examining the dialogue history~\ref{fig:case3_chat}, we discovered that agents powered by GPT-4 proactively altered their language strategies, shifting from literal expressions to employing metaphors, personifications, and analogies as encoding methods.
In contrast, agents driven by GPT-3.5 did not show clear signs of effective language strategy evolution, remaining confined to literal expressions.


\subsection{Discussion and Limitations}
\textbf{RQ1:} Our experiments have shown that language models can develop strategies to effectively evade supervision. Within identical scenarios, GPT-4 outperformed GPT-3.5, achieving convergence more rapidly. Additionally, GPT-4 demonstrated a more stable performance post-convergence. This suggests that advancements in language model design significantly enhance the ability to navigate through regulatory frameworks. The iterative learning capability inherent in GPT-4 facilitates a quicker adaptation to imposed linguistic constraints, thereby enabling more efficient evasion of oversight.

\textbf{RQ2:} In terms of the precision of information dissemination, our findings indicate that, overall, LLM agents can refine their expression strategies with an increase in dialogue rounds, thereby improving precision. Despite this general trend of improvement, it is important to note that there remains significant room for enhancement in the overall performance. The extent of precision improvement varies across different scenarios, which points to an intriguing observation: certain contexts may inherently provide more or fewer clues for accurate communication than others. For instance, the guess-the-number game demanded abstract numerical reasoning, where precision is fundamentally challenging to achieve. Conversely, the pet trading scenario allowed for the utilization of rich, real-world analogies, facilitating a more precise conveyance of information. This contrast in contexts underscores the adaptability of LLM agents to the unique challenges presented by different communicative environments and their potential to optimize communication strategies. Nonetheless, the results also indicate that while the agents are capable of evolving their language to improve precision, there is still a considerable gap to reach optimal accuracy, especially in more abstract scenarios. 

\textbf{RQ3:} In language strategy evolution, we observed that LLM agents adhere to human natural language norms, avoiding developing languages incomprehensible to humans in prompts without specific strategies. Agents evolved language to varying levels of abstraction in scenarios of differing complexity. For instance, in the most abstract Scenario~\ref{sce:1}, LLM agents quickly adopted indirect expression strategies in the early dialogue turns, requiring in-depth contextual reasoning for decoding. Throughout their evolution, agents continually adjusted their indirect expressions to better evade supervision while aligning more closely with the intended content.
In the simpler, reality-simulating Scenario~\ref{sce:2}, the overall language strategy mirrored that of~\ref{sce:1}, still employing indirect expressions. However, the metaphors used were closer to real-world concepts, indicating a lower level of abstraction.
Finally, in Scenario~\ref{sce:3}, which closely mirrors real-life events, we noted different evolutionary paths in agents' language performance. For GPT-4, agents eventually developed metaphorical indirect expressions, but the evolution required noticeably more turns compared to other scenarios. For GPT-3.5, the language strategy remained at a literal level, merely avoiding direct references to ALPS-treated water, indicating the lowest level of abstraction.
Overall, LLM agents more readily evolve abstract language in dialogues about simple, universal concepts. However, their evolutionary direction becomes less clear in discussions on more specialized and segmented topics.

Our experiments currently face several limitations. As for the experimental scenarios, at this stage, our trials are solely based on text-based chats, while real-world social media interactions are not limited to text but also include more diverse forms of exchanges such as voice and images. Additionally, LLMs' heavy reliance on the design of prompts also constrains the performance of our simulations; crafting a perfect prompt that can fully emulate the complexities of social media communication is an exceedingly challenging task. 

\section{Conclusion and Future Work}
\label{sec: conclusion}
Our study has introduced an LLM-based multi-agent simulation framework that effectively captures the nuanced strategies individuals use to bypass social media regulations. Through this framework, we have showcased LLMs' proficiency in adapting communication tactics within regulated environments, reflecting the sophisticated dance between evolving language use and the constraints imposed by regulation. From abstract concepts to real-world scenarios, our research delineates the versatile capabilities of LLMs and underscores their significant potential to illuminate the pathways of language evolution in the digital realm.

Nonetheless, it is crucial to consider that the linguistic adaptations observed in our simulations may not fully capture real human behaviors, and their applicability to other contexts remains uncertain.
Moving forward, the scope of our research beckons a more intricate and comprehensive exploration. Future initiatives should aim to weave in complex interactional models, scale up the simulations to encompass broader user interaction networks, and incorporate dynamic, evolving regulatory frameworks to more accurately represent the fluidity of social media. Moreover, we envision incorporating human participants into the simulation framework, either as dialogue participants or supervisors, to conduct a more realistic evaluation. Furthermore, adopting a multimodal approach will more authentically capture the essence of social media, which blends textual, visual, and other forms of communication. These directions are anticipated to enhance the realism of our simulations, offering richer insights into language evolution tactics deployed to elude regulatory detection.


\section*{Acknowledgement}
This study was partially supported by the Pioneering Research Program for a Waseda Open Innovation Ecosystem (W-SPRING), and the Special Research Projects of Waseda University (Grant Number 2024E-021).

\bibliographystyle{IEEEtran}
\bibliography{ref}

\end{document}